\documentstyle[11pt,epsfig,psfrag,wrapfig]{article}
\newcommand{\be}{\begin{equation}}
\newcommand{\ee}{\end{equation}}
\newcommand{\ba}{\begin{eqnarray}}
\newcommand{\ea}{\end{eqnarray}}

\newcommand{\Pnperp}{{\bf P}_{\!\!\!\perp{\mbox{\tiny N}}}}
\newcommand{\modPnperp}{P_{\!\!\!\perp\mbox{\tiny N}}}
\newcommand{\Mn}{M_{\mbox{\tiny N}}}
\newcommand{\HERMES}{{\mbox{\tiny HERMES}}}

\newcommand{\limNR}{\lim\limits_{ {\renewcommand{\arraystretch}{0.5}
	\begin{array}{c}\mbox{\tiny non}\cr\!\!\!\mbox{\tiny relativistic}
	\!\!\!\end{array}}} }
\newcommand{\di}{ {\rm d} }
\newcommand{\la}{\langle}
\newcommand{\ra}{\rangle}

\begin{document}
\title{\bf\boldmath
	Azimuthal asymmetries at CLAS: \\
	Extraction of $e^a(x)$ and prediction of $A_{UL}$}
\author{A.~V.~Efremov$^a$, K.~Goeke$^b$, P.~Schweitzer$^c$ \\
  	\footnotesize\it $^a$ 
  	Joint Institute for Nuclear Research, Dubna, 141980 Russia\\
  	\footnotesize\it $^b$ 
  	Institut f\"ur Theoretische Physik II, 
  	Ruhr-Universit\"at Bochum, D-44780 Bochum, Germany\\
  	\footnotesize\it $^c$ 
  	Dipartimento di Fisica Nucleare e Teorica, 
  	Universit\`a degli Studi di Pavia, I-27100 Pavia, Italy}
\date{}
\maketitle
%
\begin{abstract}
\noindent
First information on the chirally odd twist-3 proton distribution function 
$e^a(x)$ is extracted from the azimuthal asymmetry, $A_{LU}$, in the
electro-production  of pions from deeply inelastic scattering of
longitudinally polarized electrons off unpolarized protons, which has been 
recently measured by  CLAS collaboration.
Furthermore parameter-free predictions are made for azimuthal asymmetries,
$A_{UL}$,  from scattering of an unpolarized beam on a polarized proton target
for CLAS kinematics. 
\end{abstract}

\section{Introduction}

Experimental information on the chirally odd twist-3 proton distribution 
function $e^a(x)$ \cite{Jaffe:1991kp,Jaffe:1991ra} from deeply inelastic 
scattering (DIS)  would provide not only insights into the twist-3 nucleon 
structure.
The first moment of $e^a(x)$ is related to the pion-nucleon $\sigma$-term,
which in turn is related to the strangeness content of the nucleon.
Here one faces the so called ``$\sigma$-term puzzle''.
Results from chiral perturbation theory and the value 
$\sigma \simeq (60-80)\,{\rm MeV}$ extracted from pion-nucleon scattering 
data \cite{Koch:pu,Pavan:2001wz} imply that around $10\%$ of the nucleon mass 
is due to the strange quark.
This contrasts the fact that strange quarks carry a negligible fraction of
the nucleon momentum at say $1\,{\rm GeV}^2$, the ``typical hadronic scale'' 
for nucleon set by the nucleon mass $\Mn$.

Since $e^a(x)$ is a spin-average distribution, it can be accessed
in experiments with unpolarized nucleons. 
However, due its chiral-odd nature and twist-3 character it can enter an
observable  only in connection with another chirally odd distribution or
fragmentation  function, and with a power suppression $\Mn/Q$, where $Q$ is
the hard scale  of the process.
So one is lead to study processes at moderate $Q$, to which 
$e^a(x)$ gives the leading contribution.

An observable, where $e^a(x)$ appears as leading contribution, is the 
azimuthal asymmetry $A_{LU}$ in pion electro-production from semi-inclusive DIS
of polarized  electrons off unpolarized protons
\cite{muldt,Mulders:1996dh}\footnote{In $A_{XY}$ $X(Y)$ denotes beam (target)
 	polarization, and one should take the values $U$ for unpolarized, 
	$L$ for longitudinally polarized. We use the notation of 
	\cite{muldt,Mulders:1996dh}, with $H_1^{\perp}(z)$ normalized to 
	$\la P_{h\perp}\ra$ instead of $M_h$.}.
In this quantity $e^a(x)$ appears in
connection with the chirally and T-odd twist-2 "Collins" fragmentation
function $H_1^{\perp a}(z)$, which describes the left-right asymmetry in
fragmentation of a transversely polarized quark of flavour $a$ into a hadron
\cite{muldt,Mulders:1996dh,collins}.  In the HERMES experiment $A_{LU}$ was
found consistent with zero  within error bars \cite{hermes,hermes-pi0}.
More recently, however, the CLAS collaboration reported the measurement of a 
non-zero $A_{LU}$ in a different kinematics \cite{Avakian-talk}.

So the CLAS data \cite{Avakian-talk} allow -- under the assumption of 
factorization -- an extraction of first experimental information on $e^a(x)$ 
from DIS, provided one knows $H_1^\perp$.
First experimental indications to $H_1^\perp$ came from studies 
of $e^+e^-$-annihilation \cite{todd}. 
The HERMES data on azimuthal asymmetries $A_{UL}$ in pion electro-production 
from DIS 
\cite{hermes,hermes-pi0} provide further information on $H_1^\perp(z)$.
In these asymmetries $H_1^\perp(z)$ enters in combination with the 
chirally odd twist-2 nucleon transversity distribution $h_1^a(x)$ 
\cite{Jaffe:1991kp,Jaffe:1991ra,transversity}, the twist-3 distribution 
$h_L^a(x)$  \cite{Jaffe:1991kp,Jaffe:1991ra}, and quark transverse momentum 
weighted moments thereof \cite{Mulders:1996dh}.
In Ref.~\cite{Efremov:2001cz} $H_1^\perp(z)$ has been extracted from the 
HERMES data \cite{hermes,hermes-pi0},  using for $h_1^a(x)$ and $h_L^a(x)$
predictions from the chiral quark soliton model \cite{h1-model}
and the instanton model of the QCD-vacuum \cite{Dressler:2000hc}.

In this note we will use the information on $H_1^\perp(z)$ obtained
in Ref.~\cite{Efremov:2001cz} to extract the twist-3 distribution $e^a(x)$ 
from the CLAS data \cite{Avakian-talk}. 
Furthermore, we will predict azimuthal asymmetries $A_{UL}$ for CLAS,
which are under current study.

\section{\boldmath The twist-3 distribution function $e^a(x)$}

The twist-3 quark and antiquark distribution functions $e^q(x)$ and
$e^{\bar q}(x)$ are defined as \cite{Jaffe:1991kp,Jaffe:1991ra}
\be\label{def-e}
	e^q(x) = \frac{1}{2\Mn} \int\!\frac{\di\lambda}{2\pi}\,
	e^{i\lambda x} \la N|\bar{\psi}_a(0)\psi_a(\lambda n)|N\ra \;,\;\;\;
	e^{\bar q}(x)=e^q(-x) \ee
where the insertion of the gauge-link is understood.
The $Q^2$-evolution has been studied in 
Refs.~\cite{Balitsky:1996uh,Belitsky:1997zw,Koike:1997bs}.
In the multi-colour limit the evolution of $e^a(x)$ simplifies to a
DGLAP-type evolution -- as it does for the other two proton twist-3 
distributions $h_L^a(x)$ and $g_T^a(x)$.
The latter give a constraint on $e^a(x)$, the ``twist-3 Soffer inequality'', 
as follows from \cite{Soffer:1995ww}
\be\label{Soffer-ineq}
	e^a(x) \ge 2 |g_T^a(x)| - h_L^a(x) \;. \ee

At small $x$ it behaves as, with some constants $c_k$,
\be\label{small-x}
	e^a(x)\stackrel{x\to 0}{\longrightarrow}
	c_1\,x^{-0.04} +c_2\,\delta(x)\;.\ee
The first term follows from Regge phenomenology $e(x)\approx x^{-(\alpha+1)}$.
However the Pomeron residue is, as is known, non-spin-flip, and thus decouples
from the chirally odd $e^a(x)$.
Therefore the small $x$-behaviour of $e^a(x)$ is determined by
the lowest lying spin flip trajectory, i.e. the one with the
scalar meson $f_0(980)$.
With the usual slope $\alpha'\approx 1\,{\rm GeV}^{-2}$ this 
yields a rise like $x^{-0.04}$.
The constant $c_1$ in Eq.~(\ref{small-x}) is proportional to 
$m_q/M_N$ due to Eq.~(\ref{e-2moment}) below.
The second term in Eq.~(\ref{small-x}), the possibility of a $\delta$-function
at $x=0$, has been recently discussed in Ref.~\cite{Burkardt:2001iy}.

The first moment of $(e^u+e^d)(x)$ is related to the pion-nucleon sigma-term 
\ba
\label{e-1moment}
	\int_{-1}^1\di x\;(e^u+e^d)(x)  &=&  \frac{1}{2\Mn}\la N|\,
	\left(\bar{\psi}_u\psi_u+\bar{\psi}_d\psi_d\right)\,|N\ra
	\equiv \frac{\sigma}{m_{\rm av}\!} \;\;, \\
\label{sigma}
	\sigma &=& \frac{m_{\rm av}}{2\Mn\;}
	\la N|\,\left(\bar{\psi}_u\psi_u+\bar{\psi}_d\psi_d\right)\,|N\ra 
	= \cases{
	(64\pm 8)\,{\rm MeV} & Ref.~\cite{Koch:pu}\cr
	(79\pm 7)\,{\rm MeV} & Ref.~\cite{Pavan:2001wz}.}\;\;\;\;\ea
With the average mass of the light quarks 
$m_{\rm av}\equiv\frac12\,(m_u+m_d)\simeq 5 \,{\rm MeV}$ one obtains
\be\label{e-1moment-number}
	\int_{-1}^1\di x\;(e^u+e^d)(x) \simeq (12-16) \;.\ee
However, considering Eq.~(\ref{small-x}), this does not 
necessarily imply that $(e^u+e^d)(x)$ itself is large. 
The second moment is proportional to the number of the respective 
valence quarks $N_q$ (for proton $N_u=2$ and $N_d=1$) 
and vanishes in the chiral limit
\cite{Jaffe:1991ra}
\be\label{e-2moment}
	\int_{-1}^1\di x\;x\,e^q(x)  =   \frac{m_q}{\Mn}\; N_q \;. \ee

A model estimate for quark distributions $e^q(x)$ has been given in the
framework of the bag model \cite{Jaffe:1991ra,Signal:1997ct}.
At the estimated model scale of about $0.4\,{\rm GeV}$ 
the saturation of the ``twist-3 Soffer inequality'' Eq.~(\ref{Soffer-ineq}) 
as $e(x)= 2g_T(x) - h_L(x)$ has been observed \cite{Signal:1997ct}.
The flavour index is dropped, since the quark distributions 
of Refs.~\cite{Jaffe:1991ra,Signal:1997ct} are flavour independent.

Finally we remark that the twist-3 quark distribution $e^q(x)$ and the 
unpolarized twist-2 quark distribution $f_1^q(x)$ coincide in the
non-relativistic limit
\be\label{non-rel-limit}
	\limNR e^q(x) = \limNR f_1^q(x) = N_q\;\,\delta\!\left(x-\frac13\right)
	\;, \ee
in which the current quark mass in Eq.~(\ref{e-2moment}) is to 
be interpreted as the ``constituent quark'' mass $m_q=\frac13\,\Mn$.
The sum rule Eq.~(\ref{e-1moment-number}) is however 
strongly underestimated in this limit.

\section{The Collins fragmentation function}

The crucial ingredient for the extraction of the twist-3 distribution
function $e^a(x)$ from the azimuthal asymmetry $A_{LU}$ is the knowledge 
of $H_1^\perp(z)$.

This fragmentation function is responsible for a specific azimuthal 
asymmetry of a hadron in a jet around the axis in direction of the 
second hadron in the opposite jet 
due to transversal spin correlation of $q$ and $\bar q$.
It was the measurement of this asymmetry, 
using the DELPHI data collection \cite{todd}, 
which provided first experimental indication to $H_1^\perp$.
For the leading particles in each jet of two-jet events, 
averaged over $z$ and ${\bf k}_\perp$ and over quark flavours, 
a ``most reliable'' 
(because less sensitive to the unestimated systematic error) 
value of the analyzing power of 
$|\la H_1^{\perp}\ra / \la D_1\ra| =(6.3\pm 2.0)\%$ was found. 
Using the whole available range of the azimuthal angle 
(and thus a larger statistics) the ``more optimistic'' 
(and also more sensitive to the systematic errors) value for the analyzing
power
\be\label{apower-DELPHI}
   	\left|{\la H_1^{\perp}\ra\over\la D_1\ra}\right| = (12.5\pm 1.4)\% 
	\;\;\;\mbox{[DELPHI, extraction]} \ee
was found. 
The result Eq.~(\ref{apower-DELPHI}) refers to the scale $M_Z^2$
and to an average $z$ of $\la z\ra\simeq 0.4$ \cite{todd}.

Combining the information Eq.~(\ref{apower-DELPHI}) for $H_1^\perp$ with 
predictions for $h_1^a(x)$ and $h_L^a(x)$ from the chiral quark soliton model 
\cite{h1-model} and the instanton model of the QCD-vacuum 
\cite{Dressler:2000hc}, it was possible to describe well the HERMES data 
on the $A_{UL}$ asymmetries \cite{hermes,hermes-pi0} in a parameter-free 
approach \cite{Efremov:2001cz}.
For that a weak scale dependence of the analysing power 
Eq.~(\ref{apower-DELPHI}) had to be assumed, which however is not supported 
by studies of Sudakov suppression effects \cite{Boer:2001he}.

Furthermore, in Ref.~\cite{Efremov:2001cz}  -- assuming the model predictions 
\cite{h1-model,Dressler:2000hc} for the proton chiral odd distributions --
the $z$-dependence of the favoured pion fragmentation function $H_1^\perp(z)$
has been deduced from HERMES data \cite{hermes,hermes-pi0}.
The result refers to a scale of about $4\,{\rm GeV}^2$ 
and can be parametrized by a simple fit
\ba\label{apower-HERMES-z}
&&	H_1^\perp(z) = a\,z\,D_1(z)
	\;\;\;\mbox{with}\;\;\;	a =  0.33 \pm 0.06
	\;\;\;\mbox{for}\;\;\;	0.2 \le z \le 0.7\;, \\
&&\label{apower-HERMES-av}
	\frac{\la H_1^\perp\ra}{\la D_1\ra} = (13.8\pm 2.8)\% 
	\;\;\;\mbox{for}\;\;\; \la z\ra = 0.41 	
	\;\;\;\mbox{[HERMES, extraction]}\;, \ea
where $D_1(z)$ is the favoured unpolarized pion fragmentation function.
The errors in Eqs.~(\ref{apower-HERMES-z}) are due to experimental error 
of HERMES data \cite{hermes,hermes-pi0}.
The assumption of the predictions from \cite{h1-model,Dressler:2000hc}
introduces a model dependence, which can be viewed as a ``systematic error''
and is estimated to be around $20\%$.
The $z$-averaged value Eq.~(\ref{apower-HERMES-av}) is close to the DELPHI 
result Eq.~(\ref{apower-DELPHI}), indicating that the ratio
$\la H_1^\perp\ra/\la D_1\ra$ might indeed depend on scale only weakly. 
Note also, that HERMES data favour clearly a positive sign for the analyzing 
power.
It is noteworthy that a similar relation between the favoured fragmentation
functions $H_1^\perp(z)$ and  $D_1(z)$ (even close numerically!) was found 
in a recent model calculation \cite{Bacchetta:2002tk}.

In order to estimate the analyzing power for the CLAS experiment
we assume the result Eq.~(\ref{apower-HERMES-z}) to be valid up to $z\le 0.8$,
and to be only weakly scale dependent between HERMES $\la Q^2\ra = 4\,{\rm
GeV}^2$ and CLAS $\la Q^2\ra = 1.5\,{\rm GeV}^2$ \cite{Avakian-talk}.
Due to the particular fit Eq.~(\ref{apower-HERMES-z}),  
the analyzing power is related to the average $z$ of the experiment by 
${\la H_1^\perp\ra}/{\la D_1\ra} = a\,\la z\ra$ 
with the constant $a$ from Eq.~(\ref{apower-HERMES-z}). 
For the CLAS experiment \cite{Avakian-talk} we obtain
\be\label{apower-CLAS}
	\frac{\la H_1^\perp\ra}{\la D_1\ra} = (20 \pm 4) \% 
	\;\;\;\mbox{for}\;\;\; \la z\ra = 0.61	
	\;\;\;\mbox{[CLAS, prediction]}. \ee

\section{The azimuthal asymmetry \boldmath $A_{LU}^{\sin\phi}$}

%
	\begin{wrapfigure}{R!}{5cm}
	\vspace{-2cm}
   	\mbox{\epsfig{figure=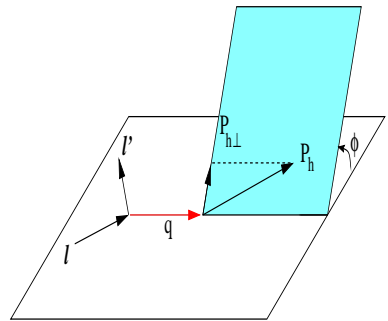,width=5cm,height=4.5cm}}
   	\caption{\footnotesize\sl
    	Kinematics of the process $ep\rightarrow e'h X$ in the lab frame.}
	\vspace{-0.2cm}
	\end{wrapfigure}
%
%
\paragraph{\boldmath $A_{LU}^{\sin\phi}$ in the CLAS experiment.}
In the CLAS experiment a longitudinally polarized $4.3\,{\rm GeV}$ electron 
beam was scattered off an unpolarized proton target. 
The cross sections $\sigma^{(\pm)}$ for the process $\vec{e}p\to e'\pi^+ X$ 
were measured in dependence of the azimuthal angle $\phi$, i.e. the angle 
between the lepton scattering plane and the plane defined by the momentum 
${\bf q}$ of the virtual photon and momentum ${\bf P}_h$ of the produced pion,
see Fig.1. 
The signs $^{(\pm)}$ refer to the longitudinal polarization of the electrons,
with $^{(+)}$ if polarization parallel to beam direction, 
and $^{(-)}$ if anti-parallel.
Let $P$ ($P_h$) be the momentum of the incoming proton (outgoing pion) and
$l$ ($l'$) the momentum of the incoming (outgoing) electron.
The relevant kinematical variables are  
center of mass energy square $s:=(P+l)^2$, 
four momentum transfer $q:= l-l'$ with $Q^2:= - q^2$,  
invariant mass of the photon-proton system $W^2:= (P+q)^2$, 
and $x$, $y$ and $z$ defined by
\be\label{notation-1}
	x := \frac{Q^2}{2Pq}	\;,\;\;\;
        y := \frac{2Pq}{s}	\;,\;\;\; 
        z := \frac{PP_h}{Pq\;} 	\;.\ee
In this notation the azimuthal asymmetry $A_{LU}^{\sin\phi}(x)$
measured by CLAS is given by
\be\label{ALU-exp}
	A_{LU}^{\sin\phi}(x) = \frac{\displaystyle
	\int\!\!\di y\,\di z\,\di\phi\,\sin\phi\,\left(
\frac{1}{S_e^{(+)}}\,\frac{\di^4\sigma^{(+)}}{\di x\,\di y\,\di z\,\di\phi}-
\frac{1}{S_e^{(-)}}\,\frac{\di^4\sigma^{(-)}}{\di x\,\di y\,\di z\,\di\phi}
\right)}{\;\;\;\;\;\;\;\displaystyle
	   \frac{1}{2}\int\!\!\di y\,\di z\,\di\phi\,\left(
   \frac{\di^4\sigma^{(+)}}{\di x\,\di y\,\di z\,\di\phi}+
   \frac{\di^4\sigma^{(-)}}{\di x\,\di y\,\di z\,\di\phi}\right)}\;\;,\ee
where $S_e^{(\pm)}$ denotes the electron polarization.
When integrating over $y$ and $z$ the experimental cuts 
have to be considered \cite{Avakian-talk}
\ba\label{exp-cuts-clas}
&	0.15\le x \le 0.4 \;,\;\;\;
  	0.5 \le y \le 0.85\;,\;\;\;
	0.5 \le z \le 0.8 \;, & \nonumber\\
&	1.0 \le Q^2/{\rm GeV}^2 \le 3.0  \;,\;\;\;
	2.0 \le W/{\rm GeV}     \le 2.6  \;.  & \ea

\paragraph{\boldmath $A_{LU}^{\sin\phi}$ in theory.}
The cross sections entering the asymmetry $A_{LU}^{\sin\phi}$ 
Eq.~(\ref{ALU-exp}) have been computed in Ref.\cite{Mulders:1996dh}
at tree-level up to order $1/Q$. 
Assuming a Gaussian distribution of quark transverse momenta one obtains
for the $A_{LU}^{\sin\phi}$ asymmetry Eq.~(\ref{ALU-exp})
\ba\label{ALU-theo}
	A_{LU}^{\sin\phi}(x)
	&=&
	\frac{1}{\la z\ra\sqrt{1+\la\Pnperp^2\ra/\la{\bf k}_{\!\perp}^2\ra}}
	\;\frac{\int\!\di y\,4y\sqrt{1-y}\,\Mn/Q^5
	       \sum_a e_a^2\,x \,e^a(x) \la H_1^{\perp a}\ra}
	       {\int\!\di y\,(1+(1-y)^2)\,/Q^4
	       \sum_b e_b^2\,f_1^b(x) \la D_1^b\ra}\;\; , \ea
where $\la\Pnperp^2\ra$ denotes the mean square transverse momentum of quarks
in the  nucleon and $\la{\bf k}_{\!\perp}^2\ra$ of the fragmenting quarks.
The latter is related to the transverse momentum of the produced pion 
by\footnote{
	Whether these relations hold exactly or only approximately,
	depends on the chosen jet selection scheme, as does the question, 
	whether $\la{\bf k}_{\!\perp}^2\ra$ is a function of $z$.
	Considering the large uncertainties on both experimental and 
	theoretical side, a discussion of jet selection scheme dependence seems 
	not appropriate here.}
$\la{\bf k}_{\!\perp}^2\ra = \la{\bf P}_{h\perp}^2\ra/\la z^2\ra$. In the 
CLAS experiment $\la P_{h\perp}\ra = 0.44\,{\rm GeV}\approx\la\modPnperp\ra$
\cite{Avakian-talk}.

Eq.~(\ref{ALU-exp}) assumes factorization to hold, 
and for that a large $Q^2$ is a necessary condition. 
Aside the general problem of factorization of $p_T$-dependent processes
there is a subtle question is whether Eq.~(\ref{ALU-exp}) can be applied 
to analyze the CLAS experiment where $\la Q^2\ra = 1.5\,{\rm GeV}^2$ 
\cite{Avakian-talk}. Here we will assume that this can be done.
This assumption will receive a certain justification, if our predictions 
on the asymmetries $A_{UL}$ (see next section) will agree well with future 
CLAS data taken at comparably low $\la Q^2\ra$.
However, one will not have a more definite answer on that, until future 
experiments performed at higher $Q^2$ will have constrained $e^a(x)$
such that a comparison between results at the different scales
-- taking $Q^2$-evolution into account -- will be possible.

%
	\begin{wrapfigure}{RD}{6cm}
	\vspace{-1.2cm}
    	\mbox{\epsfig{figure=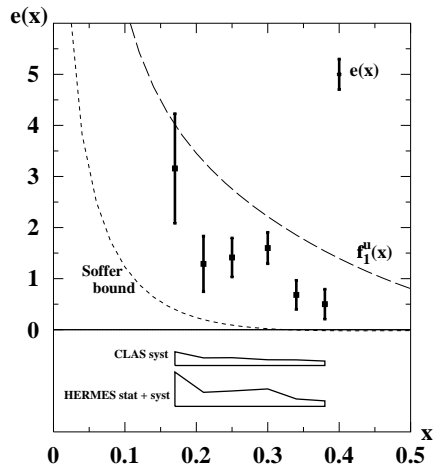,width=6cm,height=7cm}}
	\caption{\footnotesize\sl
        The flavour combination
        $e(x)\!=\!(e^u\!+\!\frac{1}{4}e^{\bar d})(x)$
	extracted from preliminary CLAS data vs. $x$ at 
	$\la Q^2\ra \!=\! 1.5\,{\rm GeV}^2$.
        The error bars are due to statistical error of the data.
        For comparison the same flavour combinations of $f_1^a(x)$ 
	and the twist-3 Soffer lower bound are shown.}
	\end{wrapfigure}
%
\paragraph{The extraction of \boldmath $e^a(x)$ from preliminary CLAS data.}
Using isospin symmetry and favoured flavour fragmentation
\be\label{eq-favored-frag}
	D_1\equiv D_1^{ u/\pi^+}\!\! = D_1^{\bar d/\pi^+}\!\! \gg
	          D_1^{ d/\pi^+}\!\! = D_1^{\bar u/\pi^+}\!\! \simeq 0 \;\ee
and the same relations for $H_1^\perp$, in the expression for the 
azimuthal asymmetry $A_{LU}^{\sin\phi}$ Eq.~(\ref{ALU-theo}), 
we see that the CLAS data yield information on the flavour combination
\be\label{def-e-flav-comb}
	e(x) \equiv e^u(x) + \frac{1}{4} e^{\bar d}(x) \;. \ee
With the estimate of the analyzing power Eq.~(\ref{apower-CLAS})
and using for $f_1^a(x)$ the parameterization of Ref.~\cite{Gluck:1994uf}
we obtain the result for $e(x)$ of Eq.~(\ref{def-e-flav-comb})
shown in Fig.~2.
For comparison the corresponding flavour combinations of the
twist-3 Soffer bound of Eq.~(\ref{Soffer-ineq})\footnote{We use 
	the ``Wandzura-Wilczek approximations''
	$g_T^a(x)=\int_x^1\di\xi\,g_1^a(\xi)/\xi$ and
	$h_L^a(x) = 2x\int_x^1\di\xi\,h_1^a(\xi)/\xi^2$. The neglect of 
	the pure twist-3 $\widetilde{h}_L^a(x)$ and $\widetilde{g}_T^a(x)$
    	is justified in the instanton QCD vacuum model 
	\cite{Dressler:2000hc,Balla:1997hf}.
	For $h_1^a(x)$ the model prediction \cite{h1-model} is used, 
	for $g_1^a(x)$ the parameterization of Ref.~\cite{Gluck:1995yr}.}
and the unpolarized distribution function $f_1^a(x)$ are plotted in Fig.~2.

Note that the uncertainties of $H_1^\perp(z)$ in Eq.~(\ref{apower-HERMES-z})
-- due to experimental error of HERMES data and theoretical assumptions in 
their analysis -- affect the overall normalization of the extracted $e(x)$.
Its $x$-dependence, however, is entirely due to the CLAS data.

The extracted $e(x)$ is clearly larger than our estimate of its twist-3 
Soffer bound Eq.~(\ref{Soffer-ineq}), about two times smaller than 
$f_1^a(x)$ at the scale of $1.5\,{\rm GeV}^2$.
The result indicates also that the large number in the sum rule 
Eq.~(\ref{e-1moment-number}) may require a significant contribution from 
the small $x$-region, which is interesting in the light of the predictions
in Eq.~(\ref{small-x}).

It is worthwhile mentioning that the bag model result for $e(x)$ of
Ref.~\cite{Signal:1997ct} (evolved according to naive power counting to 
the comparable scale of $Q^2=1\,{\rm GeV}^2$) is in qualitative agreement 
with the extracted $e(x)$.

\paragraph{\boldmath $A_{LU}^{\sin\phi}$ in the HERMES experiment.}
In the HERMES experiment the asymmetry $A_{LU}^{\sin\phi}$
has been measured with a longitudinally polarized $27.6\,{\rm GeV}$
positron beam in the kinematical range
\ba\label{exp-cuts-HERMES}
&&	0.023\le x \le 0.4 \;,\;\;\;
  	0.2 \le y \le 0.85 \;,\;\;\;
	0.2 \le z \le 0.7  \;,\nonumber\\
&&	1 \le Q^2/{\rm GeV}^2 \le 15  \;,\;\;\;
	2 \le W/{\rm GeV}     \;,\ea
and the following, consistent with zero result for the totally integrated 
asymmetries found \cite{hermes} 
\ba\label{ALU-data-HERMES}
	A_{LU}^{\sin\phi}(\pi^+)_\HERMES &=&-0.005\pm 0.008\pm0.004\nonumber\\
	A_{LU}^{\sin\phi}(\pi^-)_\HERMES &=&-0.007\pm 0.010\pm0.004
	\;\;\;\mbox{[HERMES]}\;.\ea
In order to see that the CLAS \cite{Avakian-talk} and HERMES \cite{hermes} 
data are compatible we very roughly 'parameterize' 
$e^a(x) \approx \frac12\,f_1^a(x)$ at $\la Q^2\ra = 1.5\,{\rm GeV}^2$. 
This estimate is consistent with CLAS data 
(for the flavour combination $(e^u+\frac14\,e^{\bar d})(x)$, see Fig.~2) 
and describes $e^a(x)$ sufficiently well for our purposes.
We can assume this parameterization to be valid also at the scales in 
the HERMES experiment, since evolution effects are small compared to 
the crudeness of our 'parameterization'.
This allows us to estimate $A_{LU}^{\sin\phi}(\pi^+)\approx 0.008$ and 
$A_{LU}^{\sin\phi}(\pi^-)\approx 0.007$ for HERMES kinematics, 
which is in agreement with the data in Eq.~(\ref{ALU-data-HERMES}).
We conclude that the $e^a(x)$ extracted from the CLAS experiment (Fig.~2)
is 
not in contradiction with HERMES data \cite{hermes}. 

%
\section{Predictions for \boldmath $A_{UL}$ asymmetries at CLAS}

In the HERMES experiment the azimuthal asymmetries $A_{UL}^{\sin\phi}$
and  $A_{UL}^{\sin2\phi}$ in the production of charged \cite{hermes} and
neutral \cite{hermes-pi0} pions from a proton target have been measured 
as functions of $x$ and $z$. 
For $\pi^+$ and $\pi^0$ sizeable $A_{UL}^{\sin\phi}$ asymmetries have been 
observed, while the other asymmetries have been found consistent with zero
within error bars.
In Ref.~\cite{Efremov:2001cz} the HERMES data \cite{hermes,hermes-pi0}
has been well described in a parameter-free approach, using for $H_1^\perp$
the DELPHI result \cite{todd}, see Eq.~(\ref{apower-DELPHI}), and for
proton transversity distributions the predictions from the chiral quark
soliton model \cite{h1-model} and the instanton model of the QCD vacuum 
\cite{Dressler:2000hc}. 
This approach has been used in Ref.~\cite{Efremov:2001ia} to predict 
$A_{UL}$ azimuthal asymmetries for a deuterium target.
Here we predict $A_{UL}^{\sin\phi}$ and $A_{UL}^{\sin2\phi}$ for pion 
production from a proton target for CLAS in an approach similar to
Ref.~\cite{Efremov:2001cz}, relying on the assumption that factorization 
holds at the energies of the CLAS experiment.
For the CLAS kinematics, however, the DELPHI result \cite{todd} 
for $H_1^\perp$ Eq.~(\ref{apower-DELPHI}) cannot be used, 
as it refers to a different $\la z\ra$.
Instead we use our estimate from Eq.~(\ref{apower-CLAS}).
Our predictions\footnote{For explicit expressions for 
	the azimuthal asymmetries and further details see 
	Refs.\cite{Efremov:2001cz,Efremov:2001ia,Efremov:2000za}.}
are shown in Fig.~3, for beam energies of $4.25\,{\rm GeV}$, $5.7\,{\rm GeV}$ 
and $12\,{\rm GeV}$ which are currently available or proposed for the
CLAS experiment.

Fig.~3 demonstrates that the predicted CLAS asymmetries are as large as the
asymmetries measured by HERMES \cite{hermes,hermes-pi0}.
Thus, with the high luminosity of the CLAS experiment, a precise measurement
$A_{UL}^{\sin\phi}$ and $A_{UL}^{\sin2\phi}$ for $\pi^+$ and $\pi^0$ is 
probably possible.
Moreover, the CLAS kinematics for the $12\,{\rm GeV}$ beam allows to observe 
the change of sign of the $A_{UL}^{\sin\phi}(x)$ asymmetries at $x\simeq 0.5$.
This change of sign is due to different signs of the twist-3 and twist-2 
contributions.
For the $5.7\,{\rm GeV}$ beam the $A_{UL}^{\sin\phi}(x)$ become zero close to 
the upper $x$-cut, which makes this phenomenon more difficult to observe.
For HERMES kinematics the zero of $A_{UL}^{\sin\phi}(x)$ 
lies outside the covered $x$-range and is invisible 
\cite{Efremov:2001cz,Efremov:2001ia}. 

The $A_{UL}^{\sin\phi}(x)$ asymmetries for different pions cross each other 
in a single point, see Fig.~3.
This interesting observation is due to the fact, that only two of the three 
cross sections for the production of $\pi^+$, $\pi^0$ and $\pi^-$ are 
``linearly independent'' because of isospin symmetry and favoured flavour 
fragmentation. 
Thus, if two curves cross each other in some point, the third one 
necessarily goes through this point as well.
The exact positions of this point and of the zero of $A_{UL}^{\sin\phi}(x)$ 
depend on the beam energy and move to smaller $x$ with the energy growth. 
The experimental check of this prediction, especially at COMPASS energies, 
would give an argument in favour of the handbag mechanism of the asymmetry 
with different signs of twist-2 and twist-3 contributions.

Our predictions are based on the assumption that factorization holds 
at the scales $1\,{\rm GeV}^2 \le Q^2 \le 9\,{\rm GeV}^2$ covered in
CLAS experiment \cite{Avakian-talk}. 
It will be exciting to learn from the comparison of these predictions
to future CLAS data, to which extent factorization holds.
In particular, this will give valuable indications on the
correct interpretation of the data on the $A_{LU}$ asymmetry 
and the extraction of the twist-3 distribution function
$e^a(x)$ given in the previous section.
%
%
\vspace{-0.5cm}
	\begin{figure}[t!]
	\begin{tabular}{cccccc}
	\hspace{-1.4cm} &
	\includegraphics[width=6.3cm,height=6cm]{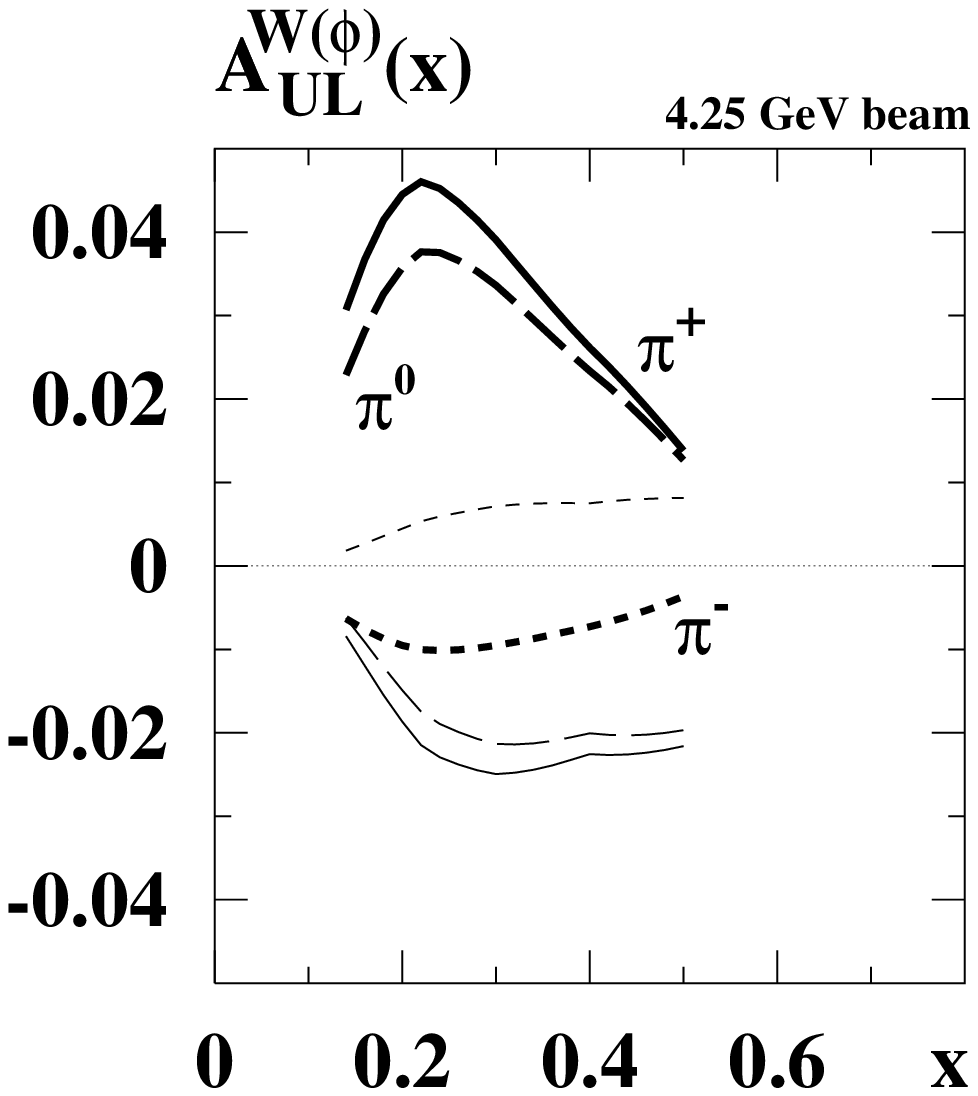} &
	\hspace{-1.4cm} &
	\includegraphics[width=6.3cm,height=6cm]{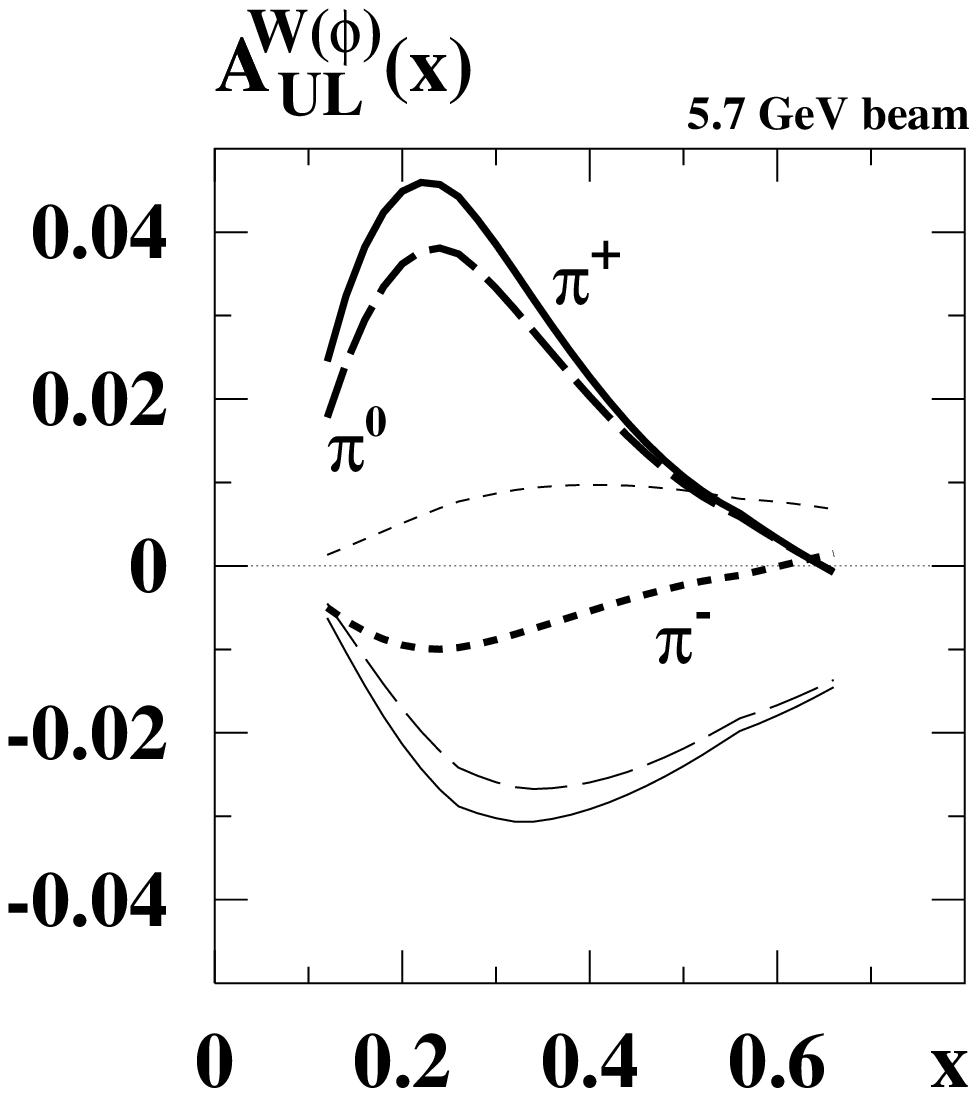} &
	\hspace{-1.4cm} &
	\includegraphics[width=6.3cm,height=6cm]{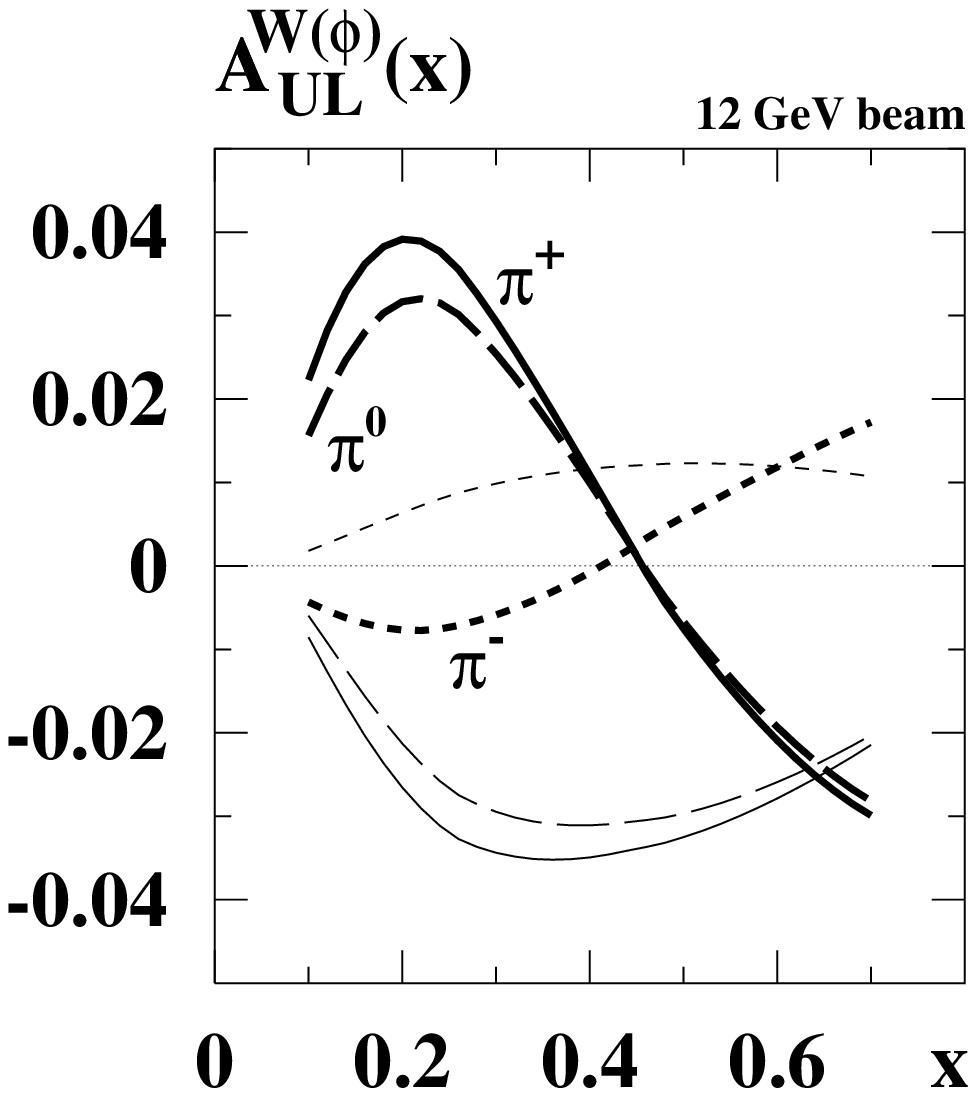}
	\end{tabular}
	\caption{\footnotesize\sl
	Predictions for azimuthal asymmetries $A_{UL}^{W(\phi)}(x)$ 
	vs. $x$ for different beam energies and the corresponding 
	kinematical cuts at CLAS. 
	The thick lines correspond to $W(\phi) = \sin\phi$,
	the thin lines correspond to $W(\phi) = \sin2\phi$. 
	Hereby solid lines refer to $\pi^+$, 
	long-dashed lines to $\pi^0$, 
	and short-dashed lines to $\pi^-$.}
	\end{figure}
%

\section{Conclusions}

We have presented the extraction of first information of the chirally 
odd proton twist-3 distribution function $e^a(x)$ from the azimuthal 
asymmetry $A_{LU}$ in $\pi^+$ electro-production from semi-inclusive DIS of
polarized  electrons off unpolarized protons, which has been recently measured
by CLAS. The flavour combination $(e^u+\frac14e^{\bar d})(x)$ extracted in the 
$x$-region $0.15\le x\le 0.4$ refers to a scale of $1.5\,{\rm GeV}^2$ 
and is sizeable -- roughly half the magnitude of the unpolarized 
distribution function at that scale.
But it is not large enough to explain the large number for the first moment 
of $(e^u+e^d)(x)$, related to the pion nucleon sigma term, by contributions
from valence $x$-regions alone.

The extraction relies on the assumption of factorization,
which might be questioned at the $Q^2$ of the CLAS experiment.
To test this assumption, we have predicted azimuthal asymmetries $A_{UL}$ 
in pion electro-production from DIS of unpolarized electrons off polarized 
protons for CLAS kinematics, which are under current study.
The predictions are based on a parameter-free approach, which has been 
shown to describe well the corresponding data from the HERMES experiment.
A successful comparison of these predictions to future CLAS data would
support the assumption of applicability of factorization at the moderate scale.

For a definite clarification of the question, whether the CLAS data has been 
interpreted here correctly, we have to wait for data from future high 
luminosity (needed to resolve the twist-3 effect) experiments
performed at scales where factorization is less questioned.
Maybe COMPASS experiment at CERN could be one of them. 
Our predictions for COMPASS will be published elsewhere.

\vspace{0.5cm}\noindent
\begin{minipage}{16cm} {\footnotesize
We would like to thank H.~Avakian for many very fruitful discussions,
and B.~Dressler for providing the evolution code.
A.~E.~is partially supported by RFBR grant 00-02-16696 and INTAS grant 00/587
and the Deutsche Forschungsgemeinschaft. This work has partly been performed
under the contract   HPRN-CT-2000-00130 of the European Commission.
}\end{minipage}


\end{document}